\documentclass[prl,twocolumn,superscriptaddress,showpacs]{revtex4}
\usepackage{graphicx,amsmath,amssymb,bm}

\begin{document}

\noindent{\bf Comment on ``{\it Ab Initio} Study of $^{40}$Ca with an
Importance-Truncated No-Core Shell Model''}

In a recent Letter~\cite{RN07}, Roth and Navr{\'a}til present an
importance-truncation scheme for the no-core shell model. The authors
claim that their truncation scheme leads to converged results for the
ground state of $^{40}$Ca.  We believe that this conclusion cannot be
drawn from the results presented in the
Letter~\cite{RN07}. Furthermore, the claimed convergence is at
variance with expectations of many-body theory. In particular,
coupled-cluster calculations indicate that a significant fraction of
the correlation energy is missing.

The truncation proposed in the Letter~\cite{RN07} is based on an
importance sampling that selects the most important particle-hole
excitations. It produces unlinked diagrams and lacks size
extensivity~\cite{Bar07}. This implies that the quality of the
truncation in large systems (such as $^{40}$Ca) cannot be judged from
its behavior in small systems (such as $^4$He or $^{16}$O). In the
absence of exact benchmark results, it is difficult to demonstrate
that this truncation yields converged results for $^{40}$Ca.  A
convincing demonstration needs to show that the truncation converges
(i) with respect to (w.r.t.) the number of states retained in the
importance sampling, (ii) w.r.t. the size of the model space (i.e., the
maximal number of excited oscillator quanta approaches $N_{\rm
max}\to\infty$), and (iii) w.r.t. the number of particle-hole
excitations. For $^{40}$Ca, the Letter~\cite{RN07} fails to provide
convincing evidence for the full convergence, in particular with 
respect to (iii).

Figure 4 of the Letter~\cite{RN07} shows the convergence for $^{40}$Ca
and the $V_{\rm UCOM}$ potential. The convergence w.r.t. the size of
the model space is not yet established due to the considerable slopes
at $N_{\rm max}=16$.  We consider the convergence w.r.t. the number of
particle-hole excitations and note that a model space of size $N_{\rm
max}$ can accommodate up to $N_{\rm max}$ particle-hole
excitations. Thus, no judgment about 4p4h excitations is possible when
$N_{\rm max}$ does not exceed the value 4 by a considerable
amount. The agreement between the 4p4h results and the exact results
up to $N_{\rm max}=4$ merely demonstrates the convergence of step (i)
in small model spaces. Beyond $N_{\rm max}=4$, no exact results are
provided to judge the quality of the 4p4h truncation. Beyond $N_{\rm
max}=8$, the 4p4h excitations become too numerous, and the description
is limited to 3p3h excitations. Clearly, the explicit demonstration of
the convergence w.r.t. the number of particle-hole excitations is
missing.
 
A second calculation for $^{40}$Ca and the $V_{{\rm low}k}$ potential
is shown in Fig. 5b of the Letter~\cite{RN07}.  This figure
demonstrates that step (i) converges for the small model space with
$N_{\rm max}=4$. It also shows that the results are converged
w.r.t. the size of the model space {\it for fixed} 3p3h
excitations. Again, the convergence w.r.t. the number of particle-hole
excitations is missing.

In the absence of a demonstrated convergence, we are interested in the
accuracy of the 3p3h result for $^{40}$Ca and the $V_{{\rm low}k}$
potential presented in the Letter~\cite{RN07}.  We have previously
performed coupled-cluster calculations~\cite{Hag07} for the same
nucleus and interaction. This method is size extensive, includes a
significant fraction of the 4p4h excitations, and is very
accurate~\cite{Bar07}. The CCSD(T) results~\cite{Hag07} indicate that
the ground state energy for $^{40}$Ca and $V_{{\rm low}k}$ is about 40
MeV lower than obtained in the Letter~\cite{RN07}. The authors of the
Letter~\cite{RN07} seem to have missed this result from our
work~\cite{Hag07}, and in their comparison for $^{16}$O, they referred
to our CCSD result but not to the more accurate CCSD(T) result.

The shortcomings of a 3p3h truncation are well known.  We give 
examples from nuclear many-body theory and quantum chemistry.  First,
$N$=$Z$ nuclei have strong $\alpha$-particle correlations which are of
4p4h character. Second, the calculation of the ground state of the
$N$=$Z$ nucleus $^{56}$Ni (See Table 1 of Ref.~\cite{Hor}) shows that
the 3p3h truncation (CISDT) is of rather poor quality compared to 4p4h
(CISDTQ) and inferior to coupled-cluster theory.  Third, quantum
chemical calculations for small molecules (see, e.g., Fig.~3 of
Ref.~\cite{Bar07}) show that the 3p3h truncation is much less accurate
than coupled-cluster theory.

In summary, the claim of a converged 3p3h truncation for $^{40}$Ca has
not been substantiated in the Letter~\cite{RN07}. This claim disagrees
with expectations from many-body theory~\cite{Bar07,Hor}, and the
reported result deviates considerably from coupled-cluster
calculations~\cite{Hag07}.

This work was supported by the U.S. Department of Energy under
the contracts DE-AC05-00OR22725, DE-FG02-96ER40963, and DE-FC02-07ER41457, 
and by the Natural Sciences and Engineering Research
Council of Canada (NSERC). TRIUMF receives federal funding via a
contribution agreement through the National Research Council of
Canada.\\

\noindent D.J.~Dean$^1$, G.~Hagen$^{1,2,3}$, M.~Hjorth-Jensen$^{3,4}$,
T.~Papenbrock$^{1,2}$, and A.~Schwenk$^5$\\
\indent $^1${\small Physics Division, Oak Ridge National Laboratory,
Oak Ridge, TN 37831, USA}\\
\indent $^2${\small Department of Physics and Astronomy, University of  
Tennessee, Knoxville, TN 37996, USA}\\
\indent $^3${\small Center of Mathematics for Applications, University of  
Oslo, N-0316 Oslo, Norway}\\
\indent $^4${\small Department of Physics, University of Oslo, 
N-0316 Oslo, Norway}\\
\indent $^5${\small TRIUMF, 4004 Wesbrook Mall, Vancouver, BC, Canada, V6T 2A3}

\end{document}